
\tolerance=1000
\newcount\eqnumber
\eqnumber=0
\def\en{\global\advance\eqnumber by 1
          \eqno(1.\the\eqnumber)}
\hsize= 13 cm
\vsize= 20 cm
\hoffset = .5 in
\voffset = .5 in
\pageno=1

\leftline{\sl Text of an invited talk given at}
\leftline{\sl The 18th International Workshop on Condensed Matter
          Theories}
\leftline{\sl Valencia, Spain, 6-10 June 1994}

\vskip 1.2 in

\centerline{\bf THEORY OF VALENCE-BOND LATTICE ON SPIN LATTICES*}
\vskip 1.5 cm
\centerline{Y. Xian}
\vskip .3 cm
\centerline{\it Department of Mathematics, UMIST}
\centerline{\it (University of Manchester Institute of Science and
                Technology)}
\centerline{\it P.O. Box 88, Manchester M60 1QD, England}
\vskip 1.5 cm

\midinsert\narrower
\noindent
{\bf Abstract.} Quantum spin-lattice systems in low dimensions exhibit
a variety of interesting zero-temperature phases, some of which show
non-classical (i.e., non-magnetic) long-range orders, such as dimer or
trimer valence-bond order. These symmetry-breaking systems with
localized valence bonds are referred to as valence-bond lattices
(VBL) in this article. A review of our systematic microscopic
formalism based on a proper set of composite operators for the ground
and excited states of the VBL systems is given. The one-dimensional
(1D) spin-$1\over2$ frustrated model is investigated in detail.
Several possible VBL systems on the 1D spin-1 chains, the 2D square
and {\it kagom\'e} lattices are also discussed. That our microscopic
theory guarantees the rotational symmetry of the VBL systems is
emphasized.
\endinsert

\vskip 2 \baselineskip
\centerline{\bf 1. INTRODUCTION}
\nobreak \vskip\baselineskip \nobreak

Spontaneous symmetry breaking (SSB) (the symmetries of a many-body
Hamiltonian not being preserved by its ground state) has always been a
fascinating phenomenon in physics. Spin-lattice systems provide ample
evidence for such SSB.  Perhaps the most well-known example is the
ferromagnetic (FM) Heisenberg models which have the classical ground
state with all spins pointing in the same direction, say, along the
$z$-axis, thereby breaking the rotational symmetry of its Hamiltonian.

The antiferromagnetic (AFM) counterparts, however, prove much more
complicated in quantum mechanics. This is well demonstrated by the
fact that even on a bipartite lattice the classical N\'eel state
consisting of two alternating spin-up and spin-down sublattices is no
longer the eigenstate of the Hamiltonian. Despite that, a number of
AFM systems show a nonzero, albeit reduced, N\'eel-like order. For
example, the 2D spin-$1\over2$ AFM Heisenberg model on the square
lattice, which has been under intensive study since the discovery of
high-temperature superconductors, is now widely believed to possess in
its ground state a N\'eel-like order which is reduced to about
two-thirds of the classical value due to quantum correlations [1,2].

In addition to the FM phase and N\'eel-like AFM phase, quantum spin
models also exhibit SSB of non-classical (i.e., non-magnetic) types in
their ground states, such as the dimerization in a spin chain with
two adjacent atoms forming a spin-singlet valence-bond (VB),
thereby breaking the chain lattice symmetry. Since the total spin
vector of an isotropic spin system is also a good quantum number,
those ground states with zero total spin vector certainly have no
classical counterparts. Any long-range order in such a many-body
ground state must be quantum mechanical in origin and the
corresponding broken symmetry is likely to be the lattice symmetry
because the rotational symmetry is preserved by such a ground state.

Working in the sector of zero total spin vector has a long history.
In fact, the seminal Bethe {\it ansatz}, which provides exact
solutions for the spin-$1\over2$ Heisenberg chain, was first proved by
Hulth\'en [3] through finite-size calculations within the framework of
resonant valence-bonds (RVB), although the term ``RVB'' was not used
then. Anderson extended this concept of RVB (originally due to
Pauline) to frustrated spin-lattice systems [4], and later to the
high-temperature superconductor materials [5]. More relevant to the
present purposes, Majumdar and Ghosh [6] found that the perfect dimer
VB configuration which breaks the lattice translational symmetry is
the exact ground state of the important 1D spin-$1\over2$ Heisenberg
chain at a particular ratio of nearest-neighbour and
next-nearest-neighbour coupling constants. In the last six years or
so, the VB basis of low-dimensional quantum spin-lattice systems has
attracted a lot of theoretical interest [7,8]. In particular, Affleck
{\it et al.} [7] discovered that a homogeneous VB configuration is the
exact ground state of a particular spin-1 Heisenberg-biquadratic
chain. This finding sheds considerable light on the well-known Haldane
conjecture [9] on the nonzero gap in the spin-1 Heisenberg chain. (The
term ``valence-bond solid'' was then used for such a VB state,
although it has no conventional symmetry breaking [10].)

An important rigorous result of quantum long-range orders is provided
by a series of spin-$s$ $SU(n)$ ($n=2s+1$) chains (or the $SU(2)$
spin-$s$ chains with Hamiltonians which project out singlet states).
This series of Hamiltonians has been solved by a mapping to the
spin-$1\over2$ $XXZ$ chain which is integrable by Bethe {\it ansatz}
[11]. By the same mapping, it has been shown that the ground state
of the $SU(n)$ model for any $n>2$ breaks the lattice symmetry with a
double degeneracy [12]. The exact values of the corresponding
dimerization order parameter have been obtained. In particular, the
dimerization order parameters for the $SU(3)$ and $SU(4)$ models are
reduced to about 42\% and 68\% respectively [12].

Although there are no exactly-known examples, it seems possible that
spontaneous trimerization, which is characterised by a sequence of
spin-singlet states formed from three adjacent spins, may occur for
some systems with integer spin quantum numbers since the trimer state
is also a rather stable configuration. Recently this possibility has
been discussed for the spin-1 Heisenberg-biquadratic chains over an
extended region of the coupling constants [13]. Furthermore, the
spontaneous dimerization or trimerization may also occur in
higher-order dimensionalities.  For example, it has been proposed that
the 2D spin-$1\over2$ frustrated Heisenberg model on the square
lattice may show a column dimer VB order over a small but nonzero
region of the coupling constants [14].  A more complicated
dimerization picture was suggested for the spin-$1\over2$ Heisenberg
model on the {\it kagom\'e} lattice [15]. The trimerization of spin-1
models on the {\it kagom\'e} lattice is also a subject to be discussed
in this article.

For convenience, the term ``valence-bond lattice'' (VBL) is used in
this article to represent collectively all those quantum spin-lattice
systems in which the simple VB configurations (i.e., dimer or trimer,
etc.) are localized with the broken lattice symmetry. One defines a
perfect VBL as a regular array of isolated simple VB configurations on
a lattice.  In general, one expects that the perfect VBL is not the
ground state of a given quantum spin-lattice Hamiltonian under
consideration. But if the system possesses a VBL long-range order in
its ground state, the perfect VBL should be a good starting point. The
quantum correlations can then be analysed on the basis of the perfect
VBL. This same strategy was employed in 1952 by Anderson [1] in his
AFM spin-wave theory, in which the quantum fluctuations from the
classical N\'eel state are described by collective motions of two sets
of bosons. The VBL systems are not unlike the quantum N\'eel-like
systems, despite the complication of their ground states being in the
sector of zero total spin vector rather than of zero total spin along
the $z$-axis.  Similar to the N\'eel-like systems, where the quantum
fluctuations are described by the spin-flip operators (i.e., spin
raising and lowering operators) with respect to the N\'eel model
state, a proper set of composite operators first developed by
Parkinson in 1979 [16] is used to describe the quantum correlations with
respect to the perfect VBL model state. A similar spin-wave theory can
then be made by a bosonization scheme for those composite operators.
But a more systematic approach is provided by a powerful microscopic
quantum many-body theory, namely the coupled-cluster method [17],
based on those composite operators themselves. The restriction in the
sector of zero total spin vector is guaranteed by a very useful and
important theorem of the CCM.

Recently, Bishop, Parkinson and Xian [18] have successfully applied
the CCM to a number of quantum spin systems, including the
spin-$1\over2$ AFM Heisenberg model on the square lattice. In their
analysis, the N\'eel state was taken as a model state for the
anisotropic-Heisenberg AFM systems.  Upon the N\'eel model state the
many-spin correlations are incorporated via a so-called correlation
operator consisting of the spin raising and lowering operators with
respect to the model state.  Within a well-defined systematic
approximation scheme amenable to computer-algebraic techniques, they
have obtained excellent results for the ground-state energy,
excitation spectra, and staggered magnetization as functions of the
anisotropy parameter. Their CCM analysis not only produces the
numerical results which are among the best estimates available today,
but also enables them to study the quantum phase transition of the
anisotropic-Heisenberg systems in an extremely systematic fashion
[19].  From these experiences, one expects that the CCM analysis
should yield similar good quantitative results for the VBL systems.

Because of its simplicity, our microscopic analysis for the
spin-$1\over2$ frustrated chains is first given. Then the same
analysis is extended to other systems, including some spin models on
the 2D square and {\it kagom\'e} lattices. The outline of this article
is as follows. Sec.~2 considers the few-body systems and introduces
the corresponding composite operators and their boson transformations.
Sec.~3 is devoted to the study of the 1D spin-$1\over2$ frustrated
model, firstly by the spin-wave approximation via a bosonization of
those composite operators, secondly by the more systematic CCM
analysis based on the composite operators themselves. Extensions of
the same analysis to the other systems, including the spin-1
Heisenberg-biquadratic chains and the some 2D models on the square
lattice and the kagom\'e lattice, are discussed in Sec.~4. A general
discussion is given in Sec.~V to conclude this article. A brief proof
of the important symmetry theorem of the CCM is given in the Appendix.

\newcount\eqnumber
\eqnumber=0
\def\en{\global\advance\eqnumber by 1
          \eqno(2.\the\eqnumber)}
\def\ket#1{{\vert{#1}\rangle}}
\def\bra#1{{\langle{#1}\vert}}
\def\vs{{\bf s}}

\vskip 2 \baselineskip
\centerline{\bf 2. FEW-ATOM SYSTEMS AND VALENCE-BONDS}
\nobreak\vskip\baselineskip\nobreak

As outlined in Sec.~1, our microscopic theory for a VBL system is
based on a set of composite operators which are defined according to
the Hilbert space of the corresponding few-atom system. Our discussion
here is restricted to the two-atom and three-atom systems.  The boson
transformation of those composite operators and the spin VB notations
are also given.

\vskip\baselineskip
\centerline{\bf 2.1 Two-Atom Systems}
\nobreak\vskip\baselineskip\nobreak

A two-atom system, each with spin $1\over2$, has four states, a
singlet and triplet states. The singlet state can be written, in
the obvious notation, as
$$ \ket0 = {1\over\sqrt2}(\ket{\uparrow,\downarrow}-
    \ket{\downarrow,\uparrow}); \en$$
and the triplet states are, respectively
$$ \ket1 =\ket{\uparrow,\uparrow},\ \
   \ket2 ={1\over\sqrt2}(\ket{\uparrow,\downarrow}
          +\ket{\downarrow,\uparrow}),\ \
   \ket3 =\ket{\downarrow,\downarrow}. \en$$
In a matrix representation, one denotes each of these four states by a
column matrix with a single nonzero element. Any operator in this
Hilbert space can then be written as a $4\times4$ matrix. Following
Parkinson [16], operator $A_{mn}$ is introduced as having only a single
non-zero element in a $(4\times4)$ matrix, namely
$\bra{m'}A_{mn}\ket{n'} = \delta_{mm'}\delta_{nn'}$. All single spin
operators can now be written in terms of these sixteen, namely
$$ \eqalignno{
s^z_1&={1\over2}(A_{02}+A_{20}+A_{11}-A_{33}),\hskip.22truein
s^z_2={1\over2}(-A_{02}-A_{20}+A_{11}-A_{33}), &(2.3a)\cr
s^-_1&={1\over\sqrt2}(A_{30}-A_{01}+A_{21}+A_{32}),\hskip.1 truein
s^-_2={1\over\sqrt2}(A_{01}-A_{30}+A_{21}+A_{32}), &(2.3b)\cr
s^+_1&={1\over\sqrt2}(A_{03}-A_{10}+A_{12}+A_{23}),\hskip.1 truein
s^+_2={1\over\sqrt2}(A_{10}-A_{03}+A_{12}+A_{23}). &(2.3c)\cr } $$
\advance\eqnumber by 1

The inverse transformations are clearly nonlinear. Therefore
$A_{mn}$ has been referred to as a composite operator [13]. In particular,
$A_{00}$ corresponds to the spin-singlet projection operator, namely
$A_{00} = 1/4 - \vs_1\cdot\vs_2$.

We notice that $A_{10}$ ($A_{30}$) is an operator which increases
(decreases) $s^z_{\rm total}\ (\equiv s^z_1+s^z_2)$ by one unit, while
$A_{20}$ leaves $s^z_{\rm total}$ unchanged; their Hermitian
conjugates (i.e., transpose matrices) have the opposite effects. Since
any of the triplet states can be generated by letting $A_{n0}$
($n=1,2,3$) operate on the singlet state $\ket0$,
$A_{n0}$ play the role of creation
operators with respect to $\ket0$; their transpose matrices
correspond to the destruction operators. Using the following algebra
$$ A_{mn}A_{kl}=A_{ml}\delta_{nk}, \en $$
which follows by definition,
it is easy to see that any $A_{mn}$ can be expressed by a product of
$A_{m0}$ and $A_{0n}$.

For a two-atom system each with spin 1, the dimension of the Hilbert
space is nine. The total number of composite operators $A_{mn}$
($m,n=0,1,2,\dots,8$) is eighty-one. Among them are eight pairs of
creation and destruction operators with respect to the singlet state.
As in the spin-$1\over2$ case, if one chooses the the singlet state
$$ \ket0={1\over\sqrt3}(\ket{1,-1}+\ket{-1,1}-\ket{0,0}),\en$$
the eight pairs of the creation and destruction
operators with respect to $\ket0$ are then denoted as $A_{n0}$ and
$A_{0n}$ $(n=1,2,\dots,8)$ respectively.

\vskip\baselineskip
\centerline{\bf 2.2 A Three-Atom System}
\nobreak\vskip\baselineskip\nobreak

Similar to the two-atom systems discussed above, in order to construct
the composite operators of a three-atom system, one should list all of
its states. Consider the spin-1 case. There are $3^3=27$ states.
Ref.~[13] lists all of them in detail. Again, the singlet state
$$ \ket0={1\over\sqrt6}(\ket{0,1,-1}+\ket{1,-1,0}+\ket{-1,0,1}
       -\ket{0,-1,1}-\ket{-1,1,0}-\ket{1,0,-1}), \en$$
is the first state, and the nine states of $s_{\rm total}=1$
($\vs_{\rm total}\equiv \vs_1+\vs_2+\vs_3$)
follow, and so on, until the last state with all three down spins.

As before, $A_{00}$ for the present case is also the spin singlet
projection operator by definition, and can be written as
$$ A_{00} = {1\over18} S_{123} (6+S_{123} -S_{123}^2), \ \ \ \
   S_{123} \equiv \vs_1\cdot\vs_2+\vs_2\cdot\vs_3+\vs_3\cdot\vs_1; \en$$
the creation and destruction operators with respect to the singlet
state $\ket0$ of Eq.~(2.6) are similarly given by $A_{n0}$ and $A_{0n}$
($n=1,2,\dots, 26$). In Ref.~13, an approximation is made by truncating
the Hilbert space from twenty-seven states to the first ten (i.e.,
restricting to the subspace of $s_{\rm total} = 0$ and 1). In this
subspace, there are only nine pairs of creation and destruction
operators which can be easily managed.

\vskip\baselineskip
\centerline{\bf 2.3 Bosonization of Composite Operators}
\nobreak\vskip\baselineskip\nobreak

A bosonization scheme for a set of operators usually starts from a
reference. The reference of the bosonization scheme (e.g.,
Holstein-Primakoff transformation) in the conventional spin-wave
theory [1] is either the spin-up state or spin-down state.  The
reference for our present purpose is clearly the singlet state $\ket0$
of the corresponding few-spin system. Therefore, the similarities
between operators $s^z$ and $A_{00}$ and between $s^+$ ($s^-$) and
$A_{n0}$ ($A_{0n}$) can be clearly seen.

While $s^z$ and $s^\pm$ obey the usual $SU(2)$ angular momentum
algebras, from Eq.~(2.4), it is easy to see that
$A_{mn}$ obey the following pseudo-spin algebra,
$$ 	[A_{mn},\ A_{kl}]=A_{ml}\delta_{nk} - A_{kn}\delta_{lm}. \en $$
Therefore, $A_{mn}$ has also been referred to as a pseudo-spin
operator. From Eq.~(2.8) one can make the following
Dyson-Mal\'eev-like transformation,
 $$ A_{00}=1-\sum_{n=1}^{d-1} a^+_na_n;\ \
    A_{n0}=a^+_nA_{00}, \ A_{0n}=a_n;\ \
     A_{mn}=a^+_ma_n, \en $$
where $m,n = 1,2,\dots, d-1$ with $d$ the Hilbert space dimensionality
of the few-atom system, and where $a_n,a^+_n$ are ($d-1$) sets of boson
operators, obeying the usual boson commutation, $[a_m,\ a_n^+]
=\delta_{mn}$.

By definition, the singlet state $\ket0$ is the vacuum state of the
bosons, namely, $a_n \ket 0 = 0,\ n=1,2,\dots,d-1$.  The physical
states correspond to the vacuum state $\ket 0$ and the $(d-1)$ states
with only one boson excited. Furthermore, as the matrix elements
between the physical and unphysical subspaces are equal to zero, the
transformation given by Eqs.~(2.9) is exact at zero temperature just
as in the case of the conventional spin-wave theory [1].

\vskip\baselineskip
\centerline{\bf 2.4 Valence Bonds}
\nobreak\vskip\baselineskip\nobreak

In the above analysis, the singlet state is always taken as the
reference with respect to which creation and destruction operators are
defined.  This is the essence of our microscopic theory for the VBL
systems.

As is well known, spin singlet states can be conveniently  represented
in terms of VB which in turn can be expressed by
Schwinger bosons [7,8]. Schwinger boson representation is given by the
following transformation,
$$ s^+=a^+b, \ \ s^-=ab^+, \ \ s^z={1\over2}(a^+a-b^+b), \en$$
where $a,a^+$ and $b,b^+$ obey the usual boson commutations. It should
be emphasized that Schwinger bosons are used here purely for the
notational purpose.  They should not be confused with the bosonization
scheme of Eq.~(2.9).  A spin-$s$ state with $s^z=m\ (-s\le m \le s)$
is written in the Schwinger representation as
$$ \ket{s,m} = {(a^+)^{s+m}\over\sqrt{(s+m)!}}
               {(b^+)^{s-m}\over\sqrt{(s-m)!}}\, \ket V, \en$$
where $\ket V$ is the vacuum state of the bosons.  A spin VB between
atoms $i$ and $j$ is defined by a number of the so-called VB operators
[8]
$$ C_{ij}^+ = a^+_ib^+_j-a^+_jb^+_i, \en$$
acting on the vacuum state $\ket V$. Using Eq.~(2.11), it is easy to
see that the spin-singlet states of Eqs.~(2.1), (2.5) and (2.6) can be
conveniently written respectively, apart from the trivial
normalization factors, as one-bond $C_{12}^+\ket V$, two-bond
$(C_{12}^+)^2\ket V$, and three-bond $C_{12}^+C_{23}^+C_{31}^+\ket V$
configurations.

\topinsert\narrower
\vskip 1.4 in
\noindent
{\bf Figure~1.} Three VB configurations: (a) spin-$1\over2$ dimer,
(b) spin-1 dimer and (c) spin-1 trimer. A single bond is defined by
$C_{ij}^+$ of Eq.~(2.12).
\vskip\baselineskip
\endinsert

A general VB configuration can be easily drawn for a spin-$s$ many-spin
system. In Fig.~1, two dimer and one trimer configurations are shown.
A many-spin ground state with zero total spin vector [$\vs_{\rm total}
\ (\equiv \sum_i \vs_i) =0$] is in general given by a linear summation of
all independent VB configurations in which each atom is linked by
$2s$ VBs [7,8]. In general, different many-spin VB states are not
orthogonal to one another. This makes working in the VB basis very
difficult.  But the ground state of some interesting quantum systems
is dominated by a particular VB state consisting of an array of
independent simple VBs such as those shown in Fig.~1.  These are the
VBL systems defined in Sec.~1.  In the following sections, a
systematic microscopic theory is developed by taking these perfect
VBLs as the reference state and by employing the creation and destruction
operators $A_{n0}$ and $A_{0n}$ with respect to this reference.

\def\cs{{\cal S}}
\def\cx{{\cal X}}
\newcount\eqnumber
\eqnumber=0
\def\en{\global\advance\eqnumber by 1
          \eqno(3.\the\eqnumber)}

\vskip 2 \baselineskip
\centerline{\bf 3. THE SPIN-$1\over2$ FRUSTRATED CHAINS}
\nobreak\vskip\baselineskip\nobreak
\centerline{\bf 3.1 Spin-Wave Theory}
\nobreak\vskip\baselineskip\nobreak

The 1D spin-$1\over2$ frustrated model is perhaps the simplest model
with spontaneous dimerization. The model consists of $N$ atoms each
with spin $1\over2$ on a chain with nearest-neighbour
and next-nearest-neighbour interactions. The Hamiltonian is
simply
$$ H=\sum_{i=1}^N(\vs_i\cdot\vs_{i+1}+J \vs_i\cdot\vs_{i+2}), \en$$
where $J$ is the coupling constant, the usual periodic boundary
condition is assumed, and even $N$ and a unit lattice spacing are also
chosen for convenience. At $J=0$, $H$ is the well-known Heisenberg
model which was solved exactly by Bethe {\it ansatz} [3]; its ground
state is gapless and has no long-range order.  At $J=1/2$, the
ground state is given by the dimer VB configuration as shown in Fig.~1
(a) with a double degeneracy [6],
 $$ \ket D = \prod_{r=1}^{N/2}\ket 0_{2r-1,2r},
      \en$$
where the notation $\ket 0_{i,j}$ represents the singlet state of the
pair given by Eq.~(2.1).  Let $r$ denote each dimer in Fig.~1 (a), and
$\vs_1(r)$ and $\vs_2(r)$ the two spins of the dimer, Eq.~(3.1) becomes
 $$ H = \sum_{r=1}^{N/2} \bigl[\vs_1(r)\cdot\vs_2(r) +
        \vs_2(r)\cdot\vs_1(r+1) + J\vs_1(r)\cdot\vs_1(r+1) +
        J\vs_2(r)\cdot\vs_2(r+1)\bigr]. \en$$
One can then express $H$ in terms of the composite operators $A_{mn}$
by Eqs.~(2.3).

As discussed in Sec.~2, since the fluctuations with respect to $\ket D$
can be described by operators $A_{n0}^r$ and $A_{0n}^r$ ($n=1,2,3$),
one can derive the equations of motion for all of these three sets of
pairs.  By employing the usual decoupling approximations and taking
$A_{00}\approx 1$, it is easy to derive the spin-wave spectra (i.e.,
eigen modes).  Parkinson [16] employed this method to obtain the
triplet spectrum for the Heisenberg model ($J=0$).

Application of bosonization scheme not only provides a more systematic
means to obtain the excitation spectra, but also allows one to study
the ground-state properties as well. By Eqs.~(2.9), one can further
express $H$ in terms of the three sets of bosons (a polynomial up to
sixth order).  Diagonalization of the quadratic parts of $H$ by the
usual Bogoliubov transformations, one can easily obtain the
ground-state energy $E_0$ and excitation spectra $\omega_q$
within the spin-wave approximation.  They are given by respectively
 $${E_0\over N}={3\over4}\int_0^\pi{dq\over\pi}
      \bigl[\sqrt{1-(1-2J)\cos2q}-1\bigr]-{3\over8}, \en $$
and
 $$	\omega_q=\sqrt{1-(1-2J)\cos2q}. \en $$
Eq.~(3.5) agrees with that of Parkinson at $J=0$ [16]. A
discussion of these results is left to the end of this section.

\vskip\baselineskip
\centerline{\bf 3.2 The Coupled-Cluster Method}
\nobreak\vskip\baselineskip

The CCM has been successfully applied to a wide range of quantum
many-body problems in both physics and quantum chemistry [17].  The
interested reader is referred to Ref.~[17] for the general formalism
of the CCM and to Ref.~[18] for its particular application to the spin
systems with the N\'eel-like order. Here its extension to the VBL
systems is considered.

\vskip\baselineskip
\leftline{\it (a). The Ground State}
\nobreak\vskip\baselineskip\nobreak

The CCM {\it ansatz} for the ground ket state is $\ket{\Psi_g} = {\rm
e}^S\ket\Phi$, where $\ket\Phi$ is the so-called model state
which is usually chosen as an uncorrelated many-body wavefunction, and
where $S$ is the many-body correlation operator consisting purely of
the creation operators with respect to $\ket\Phi$. For the VBL problem
under consideration, it is quite natural to choose the perfect VBL
state as the model state. The creation operators with respect to
this model state are clearly given by any combination of those
operators $A^r_{n0}$ with $n=1,2,\dots,d-1$, where $d$ is the
dimensionality of the corresponding few-spin system and $r$ denotes
its vector position in the VBL. Their Hermitian conjugates $A^r_{0n}$
are the corresponding destruction operators.

The Schr\"odinger equation of the ground state, after a simple
manipulation, can then be written as
$$ {\rm e}^{-S}H{\rm e}^S\ket\Phi=E_g\ket\Phi, \en$$
where $E_g$ is the ground-state energy, and where the
similarity-transformed Hamiltonian can be expressed as a series of
nested commutators, namely
$$ {\rm e}^{-S}H{\rm e}^S=H+[H,S]+{1\over2!}[[H,S],S]+\cdots, \en$$
which usually terminates at the fourth-order for most Hamiltonians
with pair-interaction potentials [17,18]. For the present case, the
series of Eq.~(3.7) terminates because Hamiltonians always contain
a finite-order polynomial of the destruction operators.

Now let us focus on the spin-$1\over2$ dimerization. The model state
is the given by the dimer state, $\ket\Phi = \ket D$. There are three
sets of creation operators, namely $A_{10}^r, A_{20}^r$ and
$A_{30}^r$. If one restricts to the sector of zero $s^z_{\rm total}\
(\equiv \sum_i^N s^z_i)$, the correlation
operator $S=\sum_{n=1}^{N/2} S_n$, with
$$ \eqalign{
  S_1&\equiv \sum_{r=1}^{N/2}\cs_r A^r_{20}, \ \
      S_2\equiv \mathop{{\sum}'}_{r,r'}^{N/2}\left[
   \cs^{(1)}_{r,r'}A^r_{10}A^{r'}_{30}
   -{1\over2!}\cs^{(2)}_{r,r'}A^r_{20}A^{r'}_{20}\right],\cr
  S_3&\equiv \mathop{{\sum}'}_{r,r',r''}^{N/2}\left[
   \cs^{(1)}_{r,r',r''}A^r_{10}A^{r'}_{30}A^{r''}_{20}
   -{1\over3!}\cs^{(2)}_{r,r',r''}A^r_{20}A^{r'}_{20}A^{r''}_{20}\right],
   \cr}  \en$$
etc. In Eq.~(3.8) the primes on the summations imply exclusion of the
terms with any pair of indices being equal.

The ground-state energy is obtained by taking the inner product of the
Schr\"odinger equation (3.6) with the model state $\ket D$ itself,
namely
$$ E_g=\bra D {\rm e}^{-S}H{\rm e}^S\ket D; \en$$
and the correlation coefficients $\{\cs_{r,r',...}\}$ in Eqs.~(3.8)
are determined by
the coupled set of equations obtained by taking inner products of
Eq.~(3.6) with states constructed from the corresponding
destruction operators, namely
$$ \bra D A^r_{02}{\rm e}^{-S}H{\rm e}^S\ket D = 0, \ \ \ \
   \forall r , \en$$
for the one-body equation; and
$$ \bra D A^r_{01}A^{r'}_{03}{\rm e}^{-S}H{\rm e}^S\ket D=0, \ \
   \bra D A^r_{02}A^{r'}_{02}{\rm e}^{-S}H{\rm e}^S\ket D = 0,
   \ \ \forall r,r'(\not=r) \en$$
for the two-body equations. The three-body equations and higher-order
many-body equations are obtained in a similar fashion.

The exact energy equation (3.9) can be straightforwardly derived as
$$ {E_g\over N}={1\over8}[(1-2J)(2b_1^{(1)}+b_1^{(2)}-a)-3], \en$$
where the lattice symmetry is used to set
accordingly $\cs_r=a,\ \cs^{(l)}_{r_1,r_2} =
\cs^{(l)}_{r_2,r_1} = b_r^{(l)},\ {\rm with\ } l=1,2\ {\rm and\ }
r=r_2-r_1$.

The exact one-body equation (3.10) can also be easily derived.  It
couples only to the two-body and three-body coefficients.
Similarly, the two-body equations (3.11) couple only to the one-body,
three-body and four-body coefficients, and so on.  One clearly needs
to employ an approximation
scheme for a practical calculation. The most common approximation of
the CCM is the SUB$n$ scheme which retains up to $n$-body correlation
operators. Here the SUB2 scheme is considered, namely $S\rightarrow
S_{{\rm SUB}2}=S_1+S_2$ and $S_n=0$ for $n\ge3$. The one-body equation
(3.10) yields an interesting solution, $a=0$, implying no one-body
correlations for the dimerization problem. Furthermore, the two-body
equations (3.11) provide a solution in which the two sets of two-body
coefficients are identical, namely
$$ b^{(1)}_r = b^{(2)}_r \equiv b_r. \en$$
We notice that the model state $\ket D$ is in the sector of
$\vs_{\rm total}=0$, and the one-body correlation
operator $S_1$ will take the state out of this sector. We also notice
that the two-body correlation operator $S_2$ commutes with
$\vs_{\rm total}$ if and only if Eq.~(3.13) is
satisfied. All these imply that the ground state in our SUB2
approximation remains in the sector of $\vs_{\rm total} = 0$
despite the fact that we started with operators in the sector of
$s^z_{\rm total}=0$. In fact, this nice property also holds at
higher-order approximations in the above CCM analysis. Appendix
provides a brief proof for a general theorem which states that the CCM
coupled equations [e.g., Eqs.~(3.10) and (3.11)] at any level of
approximations always provide at least a solution which guarantees the
symmetry of the model state if this symmetry is one of those belonging to
the model Hamiltonian. This is certainly a big advantage because it is
much more difficult to work in the sector of $\vs_{\rm total}=0$ than
of $s^z_{\rm total}=0$.

The energy equation is now reduced to
$$ {E_g\over N}={3\over8}[(1-2J)b_1-1], \en$$
and, after simplification, the two equivalent
two-body equations in the SUB2 scheme are given by
$$ {1\over2}\sum_{\rho=\pm1}(K_3\delta_{r\rho}+K_2b_r-2K_1b_{r+\rho}
   +K_1\sum_{r'\not=0}b_{r'}b_{r+\rho-r'}) = 0,\ \ \ \ r\not=0 \en$$
with $K_1\equiv 1-2J,\ K_2\equiv 4(1-2K_1b_1)$, and $K_3\equiv
K_1(1+4b_1^2)-2(1+2J)b_1$.  A simpler approximation can be made from
Eq.~(3.15), namely the SUB2-2 scheme which retains only the single
coefficient, $b_1$.  Eq.~(3.15) then reduces to
$$ 1-2J+2(3-2J)b_1-9(1-2J)b_1^2=0, \en$$
which is easily solved.  The full SUB2 equation (3.15) can also be
solved analytically by a Fourier transformation method in a similar
fashion as described in Ref.~18. Here only
the final result is given by the following self-consistency
equation for $b_1$,
$$ b_1={1\over3K_1}\left(2-{K_2\over2}{1\over2\pi}\int_{-\pi}^\pi
    dq\,\sqrt{1-k_1\cos2q+k_2\cos^22q}\right), \en$$
where the constants $k_1$ and $k_2$ are defined by
$$ k_1\equiv {1\over K_2^2}(4K_1K_2+8K_1^2b_1-4K_1^2X), \ \
   k_2\equiv {4K_1(K_1-K_3)\over K_2^2},   \en$$
and where $X \equiv \sum_{r=1}^{N/2}b_rb_{r+1}$, which can be
calculated self-consistently as $b_1$ of Eq.~(3.17). After $b_1$ is
determined as a function of $J$, the ground-state energy is obtained
by Eq.~(3.14). Again the discussion of these results is left to the end.

\vskip\baselineskip
\leftline{\it (b). The Excited States}
\nobreak\vskip\baselineskip\nobreak

The CCM {\it ansatz} for the excited state is $\ket{\Psi_e} =
X\ket{\Psi_g} = X {\rm e}^S\ket\Phi$, where $\ket{\Psi_g}$ is the
ground state as determined above and $X$ is the excitation correlation
operator consisting only of the creation operators as $S$ does. Using
the Schr\"odinger equations for the ground and excited states, one
obtains
$$ {\rm e}^{-S}[H,X]{\rm e}^S\ket\Phi =e X\ket\Phi, \ \
    e\equiv E_e-E_g. \en$$

In the so-called SUB(2,1) scheme, one retains up to two-body
correlations in $S$ and one-body correlations in $X$, namely,
$S\rightarrow S_{{\rm SUB}2} = S_1 +S_2$ and $X\rightarrow X_1$.
Therefore one writes, $X_1 = \sum_r \cx_r A_{10}^r$.
The other two one-body excitation operators are given by replacing
$A_{10}^r$ by $A_{20}^r$ and $A_{30}^r$ respectively.
The coefficient $\cx_r$
is determined by the inner product of Eq.~(3.19) with the state
$A_{10}^r\ket\Phi$. A Fourier transformation readily yields the
following excitation spectrum with a lattice momentum $q$,
$$ e_q = {K_2\over4}\sqrt{1-k_2\cos 2q + k_2\cos^2 2q}, \en$$
where the constants $K_2,k_1$ and $k_2$ are as defined before. The
other two excited states with operator $A_{20}^r$ and $A_{30}^r$ produce
the same spectrum. Therefore $e_q$ is a triplet spectrum as expected.

\vskip\baselineskip
\centerline{\bf 3.3 Discussion}
\nobreak\vskip\baselineskip\nobreak

Fig.~2 shows the results of the ground-state energy per atom as a
function of $J$ from the spin-wave approximation, and from the SUB2-2
and full SUB2 schemes of the CCM. The numerical results [20] of
Tonegawa and Harada, obtained by extrapolating the finite-size
calculations for $J<1/2$, and the exact results by Parkinson of the
$N=20$ system for $J>1/2$, are also included for comparison. At
$J=1/2$ (i.e., the Majumdar-Ghosh point) both spin-wave theory and the
CCM approximations give the exact result of $-3/8$. This is not
surprising because both take the dimer state $\ket D$ as their model
state. At $J=0$ (the Heisenberg point), spin-wave theory yields
$-0.4498$, while the SUB2-2 and full SUB2 schemes yield $-0.4268,
-0.4298$ respectively.  They all agree with the exact result of
$-0.4432$ by the Bethe ansatz [3]. But we notice that at $J=0$
spin-wave theory produces divergent results for other physical
quantities such as the dimerization parameter discussed later.
Furthermore, as can be seen from Fig.~2, the energy curve of spin-wave
theory is symmetric about $J=1/2$, while the extremely simple SUB2
scheme gives much better results for a wide range of the coupling
constant $J$ when compared with the numerical results of Ref.~[20].

\topinsert\narrower
\vskip 3.5 in
\noindent
{\bf Figure~2.} Ground-state energy per atom as a function of $J$.
Shown are the results from the spin-wave theory (dotted), the SUB2-2
scheme (short-dashed), and the full SUB2 scheme (long-dashed). The
terminating points are indicated. The numerical results from Ref.~[20]
are also included (solid).
\vskip\baselineskip
\endinsert

We notice that both spin-wave theory and the SUB2 scheme have two
terminating points $J_c^{(1),(2)}$, beyond which, namely for $J <
J_c^{(1)}$ and $J > J_c^{(2)}$, there is not real solution. For
spin-wave theory, the two points are given by $J_c^{(1)}=0$ and
$J_c^{(2)} = 1$, while $J_c^{(1)}=-0.4443$ and $J_c^{(2)}=1.591$ from
the SUB2 scheme.  The corresponding energy values of the SUB2 scheme
are $-0.5172$ and $-0.6977$ respectively. In the past we had
identified the SUB2 terminating points as the phase transition
critical points for the anisotropic-Heisenberg models [18]. This is
strongly supported by the calculations of the spin correlation
functions and order parameters within the same CCM analysis. The
following discussion of the triplet spectra of the spin-wave
excitations also supports that the two terminating points $J =
J_c^{(1),(2)}$ of the present dimerization case may again correspond
to the quantum phase transitions of the frustrated systems.

The triplet spectra of Eq.~(3.5) and Eq.~(3.20) have the same
qualitative behaviour.  Fig.~3 shows the schematic plot of the
spectrum from the SUB(2,1) scheme of the CCM at several values of $J$.
The spectrum clearly shows a nonzero gap between the two terminating
points and the gap collapses at both the terminating points. In
particular, the triplet spectrum is flat with a gap value of 1 at
$J=1/2$.  This flatness implies no coupling between pairs of spins
(dimers) at $J=1/2$ within the approximations.  Since the simple dimer
state $\ket D$ is the exact ground state at this point, the two-body
correlation can be easily included in the excitation operator $X$ and
the corresponding two-body coefficients can be determined by a simple
variational procedure similar to the well-known Feynman theory for the
excitation spectrum of the ${}^4$He superfluid [21]. Hence the
excitation operator with a lattice momentum $q$ is written as,
$$ X_q = \sum_r {\rm e}^{2iqr}A_{10}^r + \sum_{r,r'}f_q(r,r')
      A_{10}^rA_{20}^{r'}. \en$$
Taking $f_q(r,r')$ as the variational parameter to optimise the
expectation value of the Hamiltonian, it is found that the gap in the
spectrum is reduced by half at $q=0$ and $\pi$, but remains
at 1 at $q=\pi/2$ [22]. (The preliminary calculations of the similar
SUB(2,2) scheme not only yield similar results at $J=1/2$ but the
whole spectrum as a function of $J$ [22].) These results of the triplet
spectrum from such a low-order approximation seem to agree with
a more substantial calculation at $J=1/2$ by Shastry and Sutherland
[23].  They obtained the spectrum of a soliton-like excitation with a
minimum gap of 0.25 at $q=0$ and $\pi$ and a maximum gap of about 1 at
$q=\pi/2$.  Tonegawa and Harada's numerical calculations [20]
confirmed the nonzero gap at $J=1/2$ and in the nearby region.  They
predicted that the gap collapses at $J_0\approx 0.3$, while Haldane
[24], who used a fermion representation, predicted this value to be
about 1/6. Recently, $J_0$ has been estimated by the conformal field
theory to be about 0.2411 [25]. In any case, this gapless point may
correspond to a phase transition from the dimerized phase to a
critical phase similar to the Heisenberg model at $J=0$.

\topinsert\narrower
\vskip 2.8 in
\noindent
{\bf Figure~3.} Schematic plots of the triplet excitation spectrum in
the SUB(2,1) scheme at several values of $J$.
\vskip \baselineskip
\endinsert

A more intriguing situation occurs for $J>1/2$, where the triplet
spectrum of both spin-wave theory and the SUB(2,1) scheme has a
minimum at $q=\pi/2$. This certainly reminds us of the magneto-roton
excitations in the fractional quantum Hall effects [26].  As $J$
increases, the minimum (spin-roton gap) decreases and finally at
$J=J_c^{(2)}$, it collapses at $q=\pi/2$. Whether or not this suggests
a phase change in the spatial periodicity of the system from double to
four-fold, for example, is still unclear. The numerical calculations
of the spin-spin correlation function [20] certainly show a more
complicated feature for $J>1/2$. In particular, the short-range
four-fold N\'eel order
($\uparrow\uparrow\downarrow\downarrow\uparrow\uparrow
\downarrow\downarrow\cdots$) is observed
for $J>1/2$, contrast to the case of $J<1/2$ where the ordinary
short-range two-fold N\'eel order
($\uparrow\downarrow\uparrow\downarrow\cdots$) is
observed [27].  As we know, at $J=\infty$, the model Hamiltonian of
Eq.~(3.1) becomes two uncoupled Heisenberg chain with a double lattice
spacing showing a four-fold spatial periodicity. Clearly, higher-order
calculations are needed to obtain a clearer picture.

To conclude this section, it should be pointed out that the
dimerization order parameter, defined by $D\equiv
\langle\vs_{i-1}\cdot\vs_i\rangle -\langle\vs_i\cdot\vs_{i+1}\rangle$,
can be easily obtained within spin-wave theory and the SUB2 scheme of
the CCM. In spin-wave theory, for example, $D$ is found to be nonzero
in the region of $0<J<1$ and gradually diminish when $J$ moves toward
the two terminating points; but at the two terminating points
($J=0,1$), $D$ diverges to $-\infty$, implying a breakdown of
spin-wave theory. The SUB2 scheme of the CCM, however, yields
converging results even at the two terminating points, as it is the
case in our previous CCM analysis for the 1D anisotropic-Heisenberg
model [18].  Since it also involves the ground bra state which is not
manifestly the Hermitian conjugate of the ground ket state in the CCM,
our CCM analysis for the dimerization order parameter and correlation
functions will appear elsewhere.

\newcount\eqnumber
\eqnumber=0
\def\en{\global\advance\eqnumber by 1
          \eqno(4.\the\eqnumber)}

\vskip 2 \baselineskip
\centerline{\bf 4. OTHER SYSTEMS}
\nobreak\vskip\baselineskip\nobreak
\centerline{\bf 4.1 The Spin-1 Chains}
\nobreak\vskip\baselineskip\nobreak

Recently, the 1D Heisenberg-biquadratic spin-1 chain has attracted
much attention because it provides very rich and interesting
quantum phases. The model Hamiltonian is given by
$$ H = \cos\theta \sum_i \vs_i\cdot\vs_{i+1}
     + \sin\theta \sum_i (\vs_i\cdot\vs_{i+1})^2, \ \ s=1 \en$$
where the coupling between spins is parametrized by $\theta$.
The FM phase is restricted to the region of $\pi/2 \le \theta \le
5\pi/4$, and the rest is non-FM.

There are a number of exact results available at several values of
$\theta$. In particular, at $\theta = -\pi/2$, the system is exactly
known to be dimerized with a nonzero gap and the corresponding order
parameter is about 42\% of the perfect dimer state [11,12]. At
$\tan\theta = 1/3$, the ground state is given by a homogeneous VB
configuration with a nonzero energy gap but with no lattice
symmetry breaking [7]. At $\theta = \pi/4$, the model is again
integrable, the ground state clearly shows a triple spatial
periodicity and the excitation spectrum becomes gapless at the lattice
momentum $q=0$ and $2\pi/3$ [28]. Based on these exact results, one
tends to conclude that the system may show different
phases representing by the three VB states respectively, namely
the dimer state as shown in Fig.~1 (b), the trimer state in Fig.~1 (c)
and the homogeneous VB state discussed in Ref.~[7].
The expectation values of the Hamiltonian with respect to these three
trial wavefunctions can be straightforwardly obtained as
$$ {E_0\over
   N}=\cases{-{4\over3}\cos\theta+2\sin\theta,&homogeneous;\cr
  -\cos\theta+{8\over3}\sin\theta,&dimer;\cr
   -{2\over3}\cos\theta+{10\over9}\sin\theta,&trimer.\cr } \en $$
These values are shown in Fig.~4 as a function of $\theta$, together
with the numerical results from finite-size exact calculations [18].
One sees that the dimer state has lower energy than that of the
homogeneous VB state for $\theta<\tan^{-1}(-1/2) \approx -26.6^\circ$;
for larger $\theta$, however, the homogeneous VB state has lower
energy.  In particular, the homogeneous VB state is the exact ground
state at $\tan \theta=1/3$ [7]. At even larger $\theta$, it is
interesting to see that the trimer state has the lowest energy. This
occurs when $\theta>\tan^{-1}(3/4)\approx 36.9^\circ$. The lower
envelope of the three curves is in general quite close to the `exact'
results over the entire non-FM region. This crude approximation
certainly seems to give a clear picture for the three-phase diagram of
the spin-1 system, so far as the ground-state energy is concerned. Of
course, the precise locations of the boundaries between these phases
given here are not to be trusted because of the gross simplification.

\topinsert\narrower
\vskip 3.5 in
\noindent
{\bf Figure~4.} Expectation values of the 1D spin-1 Hamiltonian as a
function of $\theta$ with respect to the three simple VB states:
homogeneous (dotted), dimer (short dashed), and trimer (long dash).
Also shown are the results from finite-size exact calculations
(solid).
\endinsert

 From the above analysis, it is clear that one can extend our previous
calculations to study the dimerization of the chain around the region
of $\theta = -\pi/2$ and to study possible trimerization about $\theta
\ge \pi/4$.  Chubukov [29] applied a dimerized spin-wave theory using the
Holstein-Primakoff bosonization to the Hamiltonian of Eq.~(4.1) and
indeed he found that over an extended region, the dimerized spin-wave
excitations are stable. One certainly desires to obtain also other
physical quantities, such as the ground-state energy, dimerization
order parameter and the corresponding four-spin correlation functions,
etc. The CCM analysis described in Sec.~3 for the spin-$1\over2$ model
can certainly provide a systematic means to obtain these physical
quantities.

The possible trimerization of the spin-1 chain was discussed in
Ref.~[13] where the equations of motion were derived for the creation
and destruction operators $A_{n0}^r$ and $A_{0n}^r$ ($n=1,2,\dots,26$)
with $r$ denoting each of the trimers in Fig.~1 (c). After a
truncation in the Hilbert space, namely restricting to
$n=1,2,\dots,9$, the trimerized spin-wave spectra were obtained. The
lowest mode shows a nonzero gap associated with the trimerization, and
this gap collapses at precisely $\theta = \pi/4$ and $\theta = \pi/2$.
In particular, at $\theta =\pi/4$, the spectrum becomes gapless at
lattice momentum $q=0$ and $2\pi/3$ with a spin-wave velocity of
$3/\sqrt5 \approx 1.342$.  This spectrum compares well with the exact
result of Sutherland [28], which has a spin-wave velocity of
$\sqrt2\,\pi/3 \approx 1.481$. At $\theta = \pi/2$, where the system
is known to become FM, a constant zero spectrum was obtained.  Again,
the CCM analysis using $A_{n0}^r$ and $A_{0n}$ ($n=1,2,\dots,26$) with
the trimer model state should provide more systematic and reliable
results.

\vskip\baselineskip
\centerline{\bf 4.2 The Spin-$1\over2$ Frustrated Model on the
                     Square Lattice}
\nobreak\vskip\baselineskip\nobreak

The 2D spin-$1\over2$ frustrated Heisenberg model on the square
lattice is described by the following  Hamiltonian
$$ H = {1\over2}\sum_{i,\rho} \vs_i\cdot\vs_{i+\rho}
     + {1\over2}J\sum_{i,\rho'} \vs_i\cdot\vs_{i+\rho'},
     \ \ \ \ s={1\over2} \en$$
where $i$ runs over all lattice sites, and $\rho$ and $\rho'$ over all
nearest-neighbour and next-nearest-neighbour (diagonal) sites
respectively. Because its possible relevance to the high-temperature
superconductors, a variety of techniques has been applied to this
model [14]. One now generally believes that the system shows the
classical N\'eel-like order with the ordering
wavevector ${\bf Q}=(\pi,\pi)$
for small $J$ and the collinear magnetic order [${\bf Q}=(0,\pi)$]
for $J \approx 0.65$ or larger. Between these two phases (i.e., $0.35
< J < 0.6$), however, no magnetic order is observed. Although
there is no clear consensus on the zero-temperature structure for this
nonmagnetic region, the column dimerized phase shown in Fig.~5 has
been proposed [14]. In particular, Chubukov again applied his
dimerized spin-wave theory to the Hamiltonian of Eq.~(4.3). His
results seem to agree with the numerical calculations which suggest
that the column dimer VB state may be stable around $J=1/2$. But it
is fair to say that a more systematic approach is needed before one
can reach a definite conclusion on the dimerization of the 2D square
lattice.  It is quite straightforward to extend our CCM analysis for
the 1D spin-$1\over2$ frustrated chain described in Sec.~3 to the
present 2D case. We will report these results soon.

\topinsert\narrower
\vskip 1.1 in
\noindent
Fig.~5. A column dimer VB state for the 2D frustrated Heisenberg model
on the square lattice.
\vskip\baselineskip
\endinsert

\vskip\baselineskip
\centerline{\bf 4.3 The Spin-1 Models on the
           {\it Kagom\'e} Lattice}
\nobreak\vskip\baselineskip\nobreak

Spin models on the {\it kagom\'e} lattice are another group of
frustrated systems because the ground state of the classical Ising
model on the {\it kagom\'e} lattice has infinite degeneracy. In
addition to their intrinsic theoretical interest, some spin models on
the {\it kagom\'e} lattice may have been realized in experiments. For
example, in a layered compound Sr-Cr-Ga-O, the $s=3/2$ ${\rm Cr}^{3+}$
ions form a stack of dense {\it kagom\'e} lattices separated by more
dilute triangular lattices [30]. And the spin-$1\over2$ Heisenberg
model on the {\it kagom\'e} lattice has been proposed to explain the
interesting phenomena observed in the experiments with ${}^3{\rm He}$
atoms deposited on the graphite substrate [15].

Here the case of the spin-1 models on the {\it kagom\'e} lattice is
considered. It is useful to study the following trimerized
Hamiltonian,
$$ H(J) = \left(\sum_{\langle ij\rangle}+J\sum_{(ij)}\right)
    \vs_i\cdot\vs_j, \ \ s=1,\en$$
where, as shown in Fig.~6, $\langle ij\rangle$ denote the solid bonds
and $(ij)$ the dashed bond. The symmetric model is given by $J=1$. At
$J=0$, the perfect trimer VB configuration (i.e., solid bonds in
Fig.~6) is the exact ground state. Therefore one expects that the
trimerized spin-wave theory discussed in Sec.~4.1 for the spin-1 chain
should be a good approximation for at least small values of $J$.
The method of equation-of-motion has been applied [31] to the Hamiltonian
of Eq.~(4.4) for the operators $A_{n0}^r$ and
$A_{0n}^r$ with $n=1,2,\dots, 9$, restricting the Hilbert space of
each trimer (denoted by the new lattice vector $r$) to the first ten
states. The trimerized spin-wave spectra have been obtained as
functions of $J$.  Unfortunately, the spectra are found to be stable
only when $J \le 1/2$, and the symmetric point $J=1$ seems to be
beyond this simple spin-wave approximation.

\topinsert\narrower
\vskip 1.4 in
\centerline{Fig. 6. The spin-1 {\it kagom\'e} lattice.}
\vskip\baselineskip
\endinsert

However, similar to the 1D spin-1 case discussed earlier,
one can in general consider the spin-1 Heisenberg-biquadratic
model on the {\it kagom\'e} lattice. This model
is given by adding a quadratic term to Eq.~(4.4),
$$ H' = J'\sum_{i,\rho}(\vs_i\cdot\vs_{i+\rho})^2, \en$$
where, as before, $\rho$ denotes nearest-neighbour sites on the {\it
kagom\'e} lattice. From the experience of the 1D spin-1 chain, one expects
that the quadratic term of of Eq.~(4.5) may stabilize the trimer VB
state over an extended region of $J'$ and even at the symmetric point of
$J=1$. This work is in progress.

\vskip 2 \baselineskip
\centerline{\bf 5. CONCLUSION}
\nobreak\vskip\baselineskip\nobreak

In this article, a microscopic approach to the quantum spin systems
with an anticipated VBL long-range order is described. The perfect VBL
consisting of independent simple VBs is taken as the model state and
the corresponding composite operators first developed by Parkinson are
employed.  Two approximation schemes are developed, firstly a
spin-wave theory via bosonization transformation and secondly the more
systematic analysis within the framework of the CCM.  The general
formalism for the quantum correlations in the ground and excited
states of the VBL systems are given. In particular, the simple 1D
spin-$1\over2$ frustrated model have been investigated in detail as a
demonstration. The extensions of our approach to the spin-1
Heisenberg-biquadratic chain and to the 2D frustrated models on the
square lattice and {\it kagom\'e} lattice are also discussed. The
preliminary results presented in this article are quite promising
indeed. There is much more work to do. we wish to report our new
results in the near future.

\vskip 2 \baselineskip
\centerline{\bf ACKNOWLEDGEMENTS}
\nobreak\vskip\baselineskip\nobreak

I am grateful to R.F. Bishop for introducing me the powerful CCM and
for constant encouragement without which this work would not be
possible, and to J.B. Parkinson for providing the numerical results of
finite-size calculations prior to publication. I also wish to express
my gratitude to R. Guardiola and J. Navarro for the introduction of
the computer algebra package (REDUCE) which was used to derive those
coupled set of equations in the CCM analysis.  The partial travel
support from the European Research Office of the United States Army is
also acknowledged.

\newcount\eqnumber
\eqnumber=0
\def\en{\global\advance\eqnumber by 1
          \eqno(A.\the\eqnumber)}
\def\tH{{\rm e}^{-S}H{\rm e}^S}

\vskip 2 \baselineskip
\centerline{\bf APPENDIX}
\centerline{\bf SYMMETRY THEOREM OF THE COUPLED-CLUSTER METHOD}
\nobreak\vskip\baselineskip\nobreak

In Sec.~3.2, the SUB2 scheme of the CCM provides a solution which
preserves the rotational symmetry (i.e., $\vs_{\rm total}=0$) of the
model state $\ket D$, although one started with the operators in the
sector of $s^z_{\rm total}=0$. We notice that this is true in a
general application of the CCM. A brief proof of the following theorem
is given in this Appendix: the CCM equations at any level of
approximations always provide at least a solution which guarantees the
symmetry of the model state if this symmetry is one of those belonging
to the Hamiltonian.

Let $\Lambda$ be a symmetry operator, associated with which the model state
$\ket\Phi$ has an eigenvalue $\lambda_0$, namely $\Lambda\ket\Phi =
\lambda_0\ket\Phi$. Let Hamiltonian $H$ commutes with $\Lambda$,
$[\Lambda,H] = 0$. Therefore, one has
$$ \Lambda H\ket\Phi = \lambda_0H\ket\Phi. \en$$

(I). {\it Eigen operator representation.} Let $C_I^+$ and $C_J^+$ be
the multi-configurational creation operators with respect to
$\ket\Phi$, with the set-indices $I$ and $J$ respectively labeling the
general multi-particle cluster configurations. The corresponding
destruction operators are denoted as $C_I$ and $C_J$ respectively.
Assuming that $C^+_I$ commutes with $\Lambda$,
$$ [\Lambda,C^+_I] = 0, \ \ \ \ \forall I, \en$$
but $C^+_J$ does not. Instead, $C^+_J$ has the following commutations,
$$ [\Lambda,C^+_J] = \lambda_1'(J) C^+_J, \ \ \ \ \lambda_1'(J) \not= 0,
        \ \ \ \ \forall J. \en$$
Therefore one has for any positive integer $n$
$$ \Lambda (C^+_I)^n\ket\Phi = \lambda_0(C^+_I)^n\ket\Phi,
      \ \ \ \ \forall I, \en$$
and
$$ \Lambda C^+_J\ket\Phi = \lambda_1(J) C^+_J\ket\Phi,
      \ \ \ \ \lambda_1(J)\equiv \lambda_0+\lambda_1'(J). \en$$

The CCM correlation operator $S$ is defined as
$$ S \equiv \mathop{{\sum}'}_I\cs_IC^+_I + \sum_J\cs_JC^+_J, \en$$
where the prime implies that the identity term,
$C^+_0 \equiv {\bf 1}$, is excluded.
The correlation coefficients $\{\cs_I,\cs_J\}$
of Eq.~(A.6) are determined by the  following sets of the coupled
equations,
$$ \eqalignno{
   \bra\Phi C_I\tH\ket\Phi &= 0, \ \ \ \ \forall I\ (\not=0),
              &(A.7a)\cr
   \bra\Phi C_J\tH\ket\Phi &= 0, \ \ \ \ \forall J, &(A.7b)\cr}$$
\advance\eqnumber by 1
where the expansion of the similarity-transformed Hamiltonian
$$ \tH = H + [H,S] + {1\over2!}[[H,S],S] + \cdots, \en$$
will terminate at a finite order in $S$ if $H$ contains a finite
number of destruction operators.

We notice that the states $H\ket\Phi, [H,C^+_I]\ket\Phi,
[[H,C^+_I],C^+_{I'}]\ket\Phi, \dots$, all have the eigen value
$\lambda_0$ for $\Lambda$ by Eqs.~(A.1) and (A.4). However, from
Eq.~(A.5), the state $C^+_J\ket\Phi$ has a different eigenvalue,
$\lambda_1(J)\ (\not= \lambda_0)$. Therefore they must be orthogonal
to one another, namely
$$ \eqalignno{
   &\bra\Phi C_J H \ket\Phi = 0, \ \ \ \ \forall J&(A.9a)\cr
   &\bra\Phi C_J [H, C^+_I]\ket\Phi = 0, \ \ \ \ \forall J,I &(A.9b)\cr
   &\bra\Phi C_J [[H, C^+_I],C^+_{I'}]\ket\Phi = 0,
       \ \ \ \ \forall J,I,I' &(A.9c)\cr}$$
\advance\eqnumber by 1
etc. Using Eqs.~(A.8) and (A.9), one immediately concludes that
Eqs.~(A.7) have at least a solution given by $\cs_J = 0$ for all $J$.
If this solution is taken, the correlation operator $S$ of Eq.~(A.6)
preserves the symmetry of $\ket\Phi$.

(II). {\it Non-eigen operator representation.} let $C^+_I \equiv
(C_I^+(0),C_I^+(1),...,C_I^+(n_I-1))$ be the multi-configurational
creation operators with respect to $\ket\Phi$, with $n_I$ the
dimension of the symmetry $\Lambda$ within a set $I$. Assuming $C_I^+(n)$
do not have the commutations as in Eq.~(A.2) or Eq.~(A.3).
Let the corresponding
correlation coefficients be $\cs_I\equiv(\cs_I(0),\cs_I(1), ...,
\cs_I(n_I-1))$. The CCM correlation operator is then written as
$$ S = \mathop{{\sum}'}_I \cs_I\cdot C_I^+ \equiv
       \mathop{{\sum}'}_I [\cs_I(0)C_I^+(0) + \cs_I(1)C_I^+(1)+
     \cdots + \cs_I(n_I-1)C_I^+(n_I-1)], \en$$
where the primes again imply exclusion of the identity term,
$I = 0$.

As before, the correlation coefficients
$\{\cs_I(i),i=0,1,...,n_I-1\}$ are determined by the following set of
the coupled equations
$$ \bra\Phi C_I(i)\tH\ket\Phi = 0,\ \  i=0,1,...,n_I-1
      \ {\rm and}\ \forall I\ (\not=0). \en$$

Let $B_I^+\ [\equiv (B_I^+(0), B_I^+(1), ..., B_I^+(n_I-1))] =
T_I\cdot C^+_I$, where $T_I$ is a $c$-number $(n\times n)$
matrix and is chosen such that $B_I^+(0)$ commutes with the symmetry
operator $\Lambda$, namely
$$[\Lambda,B_I^+(0)] = 0,\ \ \ \ \forall I, \en$$
 but
$$ [\Lambda,B_I^+(j)] = \lambda_j'(I)B_I^+(j), \ \
   \lambda_j'(I) \not= 0, \ \ \ \ j=1,...,n_I-1
      \ {\rm and}\ \forall I. \en$$
Multiplying Eq.~(A.11) by the corresponding elements of the $c$-number
Hermitian matrix of $T_I$, and after a simple summation, one derives
the following equivalent equations
$$ \bra\Phi B_I(j)\tH\ket\Phi = 0, \ \ \ \ j=0,1,...,n_I-1
       \ {\rm and}\ \forall I\ (\not=0), \en$$
where $S$ can be equivalently written as
$$ S = \cs_I'\cdot B_I^+, \ \ \ \
   \cs_I' \equiv \cs_I\cdot T_I^{-1}. \en$$
According to (I), there is at least a solution to Eq.~(A.14)
hence to Eq.~(A.11) in which
$\cs_I'(j) = 0$ for all $j \not= 0$. If this solution is chosen, the
symmetry of the model state is preserved. Q.E.D.

\newcount\eqnumber
\eqnumber=0
\def\refno{\global\advance\eqnumber by 1  \the\eqnumber}

\vskip 2 \baselineskip
\centerline{\bf REFERENCES} \nobreak \vskip \baselineskip \nobreak

\item{*}This paper is dedicated to my mother, Hui-E Huang (1934-1993),
a simple country woman from Hainan Island, China, who devoted her
whole life to the well-beings of her husband, four children and three
grandchildren.
\item{\refno.}P.W. Anderson, Phys. Rev. {\bf 86}, 694 (1952);
T. Oguchi, {\it ibid.} {\bf 117}, 117 (1960).
\item{\refno.}E. Manousaki, Rev. Mod. Phys. {\bf 63}, 1 (1991); T.
Barnes, Int. J. Mod. Phys. C {\bf 2}, 659 (1991).
\item{\refno.}H.A. Bethe, Z. Phys. {\bf 71}, 205 (1931); L. Hulth\'en,
Ark. Mat. Astron. Fys. A {\bf 26}, No. 11 (1938).
\item{\refno.}P.W. Anderson, Mater. Res. Bull. {\bf 8}, 153 (1973).
\item{\refno.}P.W. Anderson, Science {\bf 235}, 1196 (1987).
\item{\refno.}C.K. Majumdar and D.K. Ghosh, J. Phys. C {\bf 3}, 91
(1970); J. Math. Phys. {\bf 10}, 1388, 1399 (1969).
\item{\refno.}I. Affleck, T. Kennedy, E. Lieb, and H. Tasaki, Phys.
Rev. Lett. {\bf 59}, 799 (1987).
\item{\refno.}D.P. Arovas, A. Auerbach, and F.D.M. Haldane, Phys. Rev.
Lett. {\bf 60}, 531 (1988).
\item{\refno.}F.D.M. Haldane, Phys. Lett. {\bf 93A}, 464 (1983); Phys.
Rev. Lett. {\bf 50}, 1153 (1983).
\item{\refno.}There is, however, a hidden symmetry breaking in the
valence-bond solid. See K. Rommels and M. den. Nijs, Phys. Rev. Lett.
{\bf 59}, 2578 (1987).
\item{\refno.}J.B. Parkinson, J. Phys. C {\bf 20}, 21 (1988); M.N.
Barber and M.T. Batchelor, Phys. Rev. B {\bf 40}, 4621 (1989); A.
Kl\"umper, Europhys. Lett. {\bf 9}, 815 (1989); I. Affleck, J. Phys.:
Condens. Matter {\bf 2}, 405 (1990).
\item{\refno.}Y. Xian, Phys. Lett. A {\bf 183}, 437 (1993).
\item{\refno.}Y. Xian, J. Phys.: Condens. Matter {\bf 5}, 7489 (1993).
\item{\refno.}R.R.P. Singh and R. Narayanan, Phys. Rev. Lett. {\bf
65}, 1072 (1990); A.V. Chubukov and Th. Jolicoeur, Phys. Rev. B {\bf 44},
12050 (1991); N. Read and S. Sachdev, Phys. Rev. Lett. {\bf 66}, 1773
(1991); H.J. Schulz, T.A.L. Ziman, and D. Poilblanc, preprint (1994).
\item{\refno.}V. Elser, Phys. Rev. Lett. {\bf 62}, 2405 (1989); J.B.
Marston and C. Zeng, J. Appl. Phys. {\bf 69}, 5692 (1991).
\item{\refno.}J.B. Parkinson, J. Phys. C {\bf12}, 2873 (1979).
\item{\refno.}R.F. Bishop and H. K\"ummel, Phys. Today {\bf 40(3)},
52 (1987); R.F. Bishop, Theo. Chim. Acta {\bf 80}, 95 (1991).
\item{\refno.}R.F. Bishop, J.B. Parkinson, and Y. Xian, Phys. Rev. B
{\bf 43}, 13782 (1991); {\bf 44}, 9425 (1991); J. Phys.: Condens.
Matter {\bf 4}, 5783 (1992); J. Phys.: Condens. Matter {\bf 5}, 9169
(1993).
\item{\refno.}R.F. Bishop, R.G. Hale and Y. Xian, UMIST preprint
   (1994).
\item{\refno.}T. Tonegawa and I. Harada, J. Phys. Soc. Japan {\bf
56}, 2153 (1987); and courtesy of J.B. Parkinson.
\item{\refno.}R.P. Feynman and M. Cohen, Phys. Rev. {\bf 102}, 1189
(1956); H.W. Jackson and E. Feenberg, Rev. Mod. Phys. {\bf 34}, 686
(1962).
\item{\refno.}Y. Xian, unpublished.
\item{\refno.}S.S. Shastry and B. Sutherland, Phys. Rev. Lett. {\bf
47}, 964 (1981); W.J. Caspers, K.M. Emmett, and W. Magnus, J. Phys. A
{\bf 17}, 2697 (1984).
\item{\refno.}F.D.M. Haldane, Phys. Rev. B {\bf 25}, 4925 (1982).
\item{\refno.}K. Okamoto and K. Nomura, Phys. Lett. A {\bf 169}, 433
(1992).
\item{\refno.}S.M. Girvin, A.H. MacDonald, and P.M. Platzman, Phys.
Rev. Lett. {\bf 54}, 581 (1985); S.M. Girvin, in {\it Quantum Hall
Effects} (ed. R.E. Prange and S.M. Girvin), Springer-Verlag, New York,
1987, p.~353.
\item{\refno.}I am grateful to C. Zeng for pointing out this to me.
\item{\refno.}B. Sutherland, Phys. Rev. B {\bf 12}, 3795 (1975).
\item{\refno.}A.V. Chubukov, Phys. Rev. B {\bf 43}, 3337 (1991).
\item{\refno.}A.P. Ramiresz, G.P. Espinosa, and A.S. Cooper, Phys.
Rev. B {\bf 45}, 2505 (1992), and references there in.
\item{\refno.}Y. Xian, unpublished.

\bye